\documentclass[conference,a4paper]{APSIPA2026}
\usepackage{amsmath}
\usepackage{graphicx}
\usepackage{multirow}
\usepackage{threeparttable}
\usepackage[backend=biber,style=ieee,]{biblatex}
\usepackage{amsmath,amssymb,amsfonts}

\usepackage{xcolor}
\usepackage{booktabs} 

\newcommand{\SON}{\emph{sonata}}
\usepackage{balance}  
\usepackage{geometry}
\usepackage{fancyhdr}

\fancypagestyle{firststyle}{
  \fancyhf{}
  \fancyhead[C]{© 2026 IEEE. Personal use of this material is permitted.
Permission from IEEE must be obtained for all other uses.}
}

\begin{document}

\title{Sonalyzer-Moz: A Framework for Analyzing the Structure of Mozart’s Sonata Form}

\author{
\authorblockN{
 Jing Zhao\authorrefmark{1},
KokSheik Wong\authorrefmark{1},
Vishnu Monn Baskaran\authorrefmark{1},
Kiki Adhinugraha\authorrefmark{2},
 David Taniar\authorrefmark{3}
}

\authorblockA{
\authorrefmark{1}School of Information Technology, Monash University Malaysia, Malaysia \\
E-mails: \{jing.zhao, wong.koksheik, vishnu.monn\}@monash.edu}

\authorblockA{
\authorrefmark{2}Department of Computer Science and Information Technology, La Trobe University, Australia \\
E-mail: k.adhinugraha@latrobe.edu.au }

\authorblockA{
\authorrefmark{3}Faculty of Information Technology, Monash University, Australia \\
E-mail: david.taniar@monash.edu  }
}

\maketitle
\thispagestyle{firststyle}
\pagestyle{empty}

\begin{abstract}
The \emph{sonata} form is a musically rich and hierarchically structured form that poses significant challenges for automatic analysis.
While music structure analysis has seen strides of progress in recent years, \emph{sonata} form analysis remains in its early stages.
This is largely due to the time-consuming and high barrier of the music background requirement for annotating classical music structures.
To advance research in this area, we curated \emph{SoSA-Moz}, the first large-scale dataset featuring comprehensive \textbf{hierarchical structure} annotations. This work establishes a foundation for systematic \textbf{sonata form analysis}. 
Leveraging this newly contributed resource, we further propose \emph{Sonalyzer-Moz}, a baseline model specifically designed for investigating complex \emph{sonata} structures. This framework integrates feature aggregation with sequential modeling. Experiment results show that \emph{Sonalyzer-Moz} is capable of identifying the components' boundaries of the \emph{structure level} (i.e., Exposition, Development, and Recapitulation) that are critical to understanding \emph{sonata} form. Therefore, this method demonstrates, for the first time, the effectiveness of automatic \emph{structure level} analysis of \emph{sonata} form, and provides a robust baseline for future research in the automatic understanding of \emph{sonata} form while advancing the study of classical music structure analysis.
\end{abstract}

\section{Introduction\label{sec:intro}}
With the rapid development of neural networks, music structure analysis has advanced significantly, supporting not only music education but also a wide range of applications, such as structure-based music generation~\cite{wu2025melody,dai2021controllable}, music transcription~\cite{agrawal2021structure}, music recommendation~\cite{roy2020imusic} and more.
Despite the expressive richness of \emph{sonata} form, its complex and hierarchical structure has hindered the creation of large-scale datasets suitable for neural network approaches~\cite{zhao2023computational}.
As a result, the capabilities of neural networks have yet to be fully harnessed in the context of \emph{sonata} form analysis.

To address this challenge, we curate a large-scale audio-based dataset, namely the \emph{SoSA-Moz} dataset, specifically designed for \emph{sonata} form analysis. This dataset consists of 35 unique Mozart pieces, each with multiple performance versions, resulting in a total of 582 recordings. 
Furthermore, we provide hierarchical structure annotations for \emph{sonata} form, encompassing both the \emph{structure level} (upper level, consisting of Exposition \emph{E}, Development \emph{D}, and Recapitulation \emph{R}) and the \emph{thematic function} level (lower level, consisting of subjects). To the best of our knowledge, very few datasets to date offer annotations at the \emph{sonata} form \emph{structure level}. While the recently released BPSD dataset provides structural annotations through its publicly accessible repository, it was primarily developed for cross-version studies. To support this focus, BPSD manually edited (e.g., cutting and copying) the structures of different performance versions (e.g., converting EDR to EDRDR due to repeats), a modification that renders the data unsuitable for authentic structural analysis. Consequently, the proposed \emph{SoSA-Moz} dataset stands as the first large-scale, structurally annotated dataset specifically viable for \emph{sonata} form structure analysis.

Based on the curated \emph{SoSA-Moz} dataset, we propose the \emph{Sonalyzer-Moz} framework to analyze the structure of \emph{sonata} form, with a particular focus on the \emph{structure level}. 
However, to the best of our knowledge, no deep learning–based methods have been specifically developed for the analysis of classical music form structures. Consequently, we evaluate our approach against representative state-of-the-art methods originally designed for popular music structure analysis. Although we have fairly adjusted their parameters and retrained these models, it must be emphasized that the hierarchical structural characteristics of popular music differ substantially from those of the \emph{sonata} form. Therefore, these existing methods exhibit inherent and predictable limitations.

\begin{figure*}[!t]
\centering
\includegraphics[width=.75\textwidth]{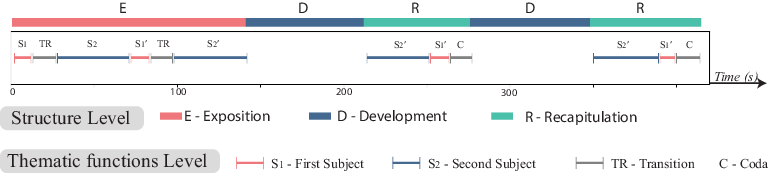}
\vspace{-3mm}

\caption{Structural visualization of the \emph{sonata} form of Mozart’s Sonata K.311, first movement.\label{fig:SonataStructure}
}
\vspace{-5mm}
\end{figure*}

Experimental results show that the proposed \emph{Sonalyzer-Moz} approach achieves 76.24\% in F1 score with a $\pm3$ second tolerance (hereinafter referred to as HR3F) on the structure level. This result is higher than that of the best retrained method originally developed for popular music. Therefore, \emph{Sonalyzer-Moz} can serve as an effective baseline method for \emph{sonata} form analysis. The code and dataset are publicly available\protect\footnotemark[1].

\footnotetext[1]{The source code, annotations, metadata, pretrained model,
and preprocessed input features are publicly available on
Figshare at
\url{https://doi.org/10.6084/m9.figshare.31745185}. }

\section{Preliminary: Sonata form\label{sec:preliminary}}

The \SON~form is a large-scale musical structure~\cite{hepokoski2006elements}. 
It is important to note that the ``Sonata music" and the ``Sonata form" are different entities. 
In all movements of Sonata music, only certain movements are in \emph{sonata} form, for example, the first or last movements~\cite{morris1935structure}. Additionally, many symphonies, particularly those from the Classical and Romantic periods, are composed in \emph{sonata} form.
Our work here focuses exclusively on movements that are in the \SON~form.

A visualization of a music piece in the \SON~form is shown in Fig.~\ref{fig:SonataStructure}.
Typically, the \SON~form consists of three main components: \emph{exposition} $E$, \emph{development} $D$, and \emph{recapitulation} $R$, which are collectively referred to as the \emph{structure level}.
In addition, the \emph{exposition} $E$ (visualized in pink) follows a complex internal organization known as the \emph{thematic function level}~\cite{hepokoski2006elements,laitz2012complete}.
In most cases, $E$ consists mainly of two subjects, namely the first subject $S_1$ and secondary subject $S_2$~\cite{morris1935structure}.
Here, the harmonic contrast between $S_1$ and $S_2$ is often used to highlight tension and emotional conflict~\cite{laitz2012complete}. 
A transition section $TR$ may appear between $S_1$ and $S_2$ to enhance coherence and facilitate modulation. 
Furthermore, a coda $C$ may be included to conclude the exposition.

After $E$, an independent component $D$ will be presented~\cite{laitz2012complete}.
In other words, the melody in $D$ could be a new material, developed from the subjects in $E$~\cite{morris1935structure, laitz2012complete}.
Therefore, $D$ is relatively free without any form constrains~\cite{laitz2012complete,spring2013musical}.
Consequently, the unpredictable nature of $D$ presents a unique challenge for analyzing \emph{sonata} form.
The last component in the \SON~form is $R$. 
It will partially recap $E$ using a similar set of components.
This component performs as a \emph{return} function, ensuring the integrity and unity of the music after undergoing the controversy and free development in $E$ and $D$, respectively~\cite{spring2013musical}. 
To achieve this function, the transition, coda, subjects may change~\cite{laitz2012complete,spring2013musical}. 
Fig.~\ref{fig:SonataStructure} illustrates an interesting case in which the occurrence order of $S_1$ and $S_2$ is reversed in section $R$. Without such a reversal, researcher could approach the analysis by detecting repeated features. However, this reversal poses additional challenges for the \SON~form analysis.

\section{Related works}\label{sec:related_work}

\subsection{Related dataset}

To date, large-scale classical music datasets typically provide only coarse-grained form labels~\cite{burger2024direct,chawin2021sliding}, such as \emph{rondo} label, \emph{sonata} label, etc.
Datasets with detailed structural annotations remain extremely scarce.
In 2024, the BPSD dataset \cite{zeitler2024bpsd} was introduced, containing 32 unique pieces in \emph{sonata} form by Beethoven, where each piece has 11 different performance versions with structural annotations. 
However, out of the 11 different versions available for each unique piece, only four (4) versions per piece are freely accessible for research purposes. 
Consequently, the dataset remains limited for data-driven methods. More importantly, although the BPSD dataset provides structural annotations, these labels are not suitable for the field of structural analysis. This is because different performances of a \emph{sonata} form piece may exhibit different performance structures. For example, a piece can be performed with an EDR structure or an EDRDR structure (i.e., including repeats). Such variations are common in \emph{sonata} form and are not considered performance errors~\cite{zeitler2024bpsd}. However, since the BPSD dataset was primarily designed for cross-version studies, the structures of different performance versions were manually edited, such as cutting and copying components, to enable alignment across versions. These modifications render the dataset unsuitable for authentic structural analysis. Consequently, a large-scale \emph{sonata} form dataset with reliable structural annotations is urgently needed for further research.

\subsection{Music Structure Analysis Methods}

To the best of our knowledge, there is currently limited work on \emph{sonata} form structure analysis. 
In the past decade, based on symbolic music files, Bigo et al.~\cite{bigo2017sketching} proposed a non-deep learning approach, using the Viterbi algorithm, to estimate the structure of Mozart’s string quartets in \emph{sonata} form.
According to their reported qualitative results, many structural components were incorrectly identified, including  
components being ``not found", large errors in segment length, and incorrect detection of segment boundaries.
In other words, even when using symbolic data that contain exact musical note information, there are no effective methods for analyzing \emph{sonata} form. Therefore, analyzing \emph{sonata} form remains a challenging task.
Furthermore, symbolic files present limitations in terms of scalability for classical music datasets because they are rigid, unlike audio recordings that can capture variations or differences in performance characteristics such as repeated sections, playing techniques, and instrument features~\cite{zhao2024music}.
Given these constraints, we anticipate that the combination of our large-scale audio-based dataset and deep learning neural network methods will create new opportunities for \emph{sonata} form analysis. 

In fact, many advanced methods have already been developed for the analysis of popular music structures, such as  AllInOne~\cite{kim2023all}, SongFormer~\cite{hao2025songformer}, etc. However, classical music generally exhibits more complex hierarchical characteristics as well as richer variations and content. Consequently, we assume that methods designed for popular music are unlikely to perform well on classical music without considering the structural theories specific to its forms. Nevertheless, in this study, we still compare our approach with representative state-of-the-art methods for popular music, in order to validate our assumptions and provide a discussion from the perspective of methodological design.

\begin{figure*}[t!]
\centering
\includegraphics[width=1\textwidth]{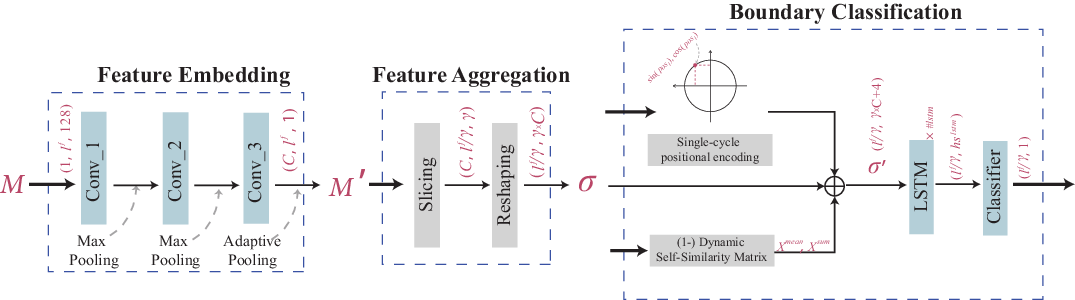}
\vspace{-5mm}
\caption{Visualization of the \emph{Sonalyzer-Moz} framework.\label{fig:Sonalyze}
}
\vspace{-5mm}
\end{figure*}

\section{The \emph{SoSA-Moz} dataset}\label{sec:dataset}
\begin{table}[t!]
\caption{Meta information of the \emph{SoSA-Moz} dataset.\label{table:SOSA-MOZ}}
\vspace{-3mm}

\centering
\resizebox{0.35\textwidth}{!}{
\begin{tabular}{lc} \toprule \midrule
\# of pieces    &   582               \\
\# of unique pieces      &   35                \\
Instrument       &   Piano \& Strings     \\
Composer   &   W.A. Mozart             \\\midrule \bottomrule
\end{tabular}
}
\vspace{-3mm}

\end{table}

The \emph{SoSA-Moz}\protect\footnotemark[1] (\emph{sonata} Form Structure Analysis For Mozart) is a multi-version dataset. It provides structural annotations for 582 \emph{sonata} form movements by Mozart, comprising 35 unique works, each with $\sim$16 different performance versions. 
An overview of the dataset is provided in Table~\ref{table:SOSA-MOZ}.

The \emph{SoSA-Moz} dataset curation process included identifying multiple performance versions on YouTube, downloading the selected recordings for internal research processing, and conducting final quality filtering.
Following this, the dataset was annotated by three annotators, each with formal training in classical music performance. To ensure annotation reliability, each piece was independently annotated by two different annotators as part of a cross-validation procedure.
Furthermore, to ensure the authoritativeness of the annotations, all labeling was strictly guided by established music analysis literature~\cite{marks1921sonata, flothuis1998mozarts}. 
Consequently, the annotation process was conducted objectively, without reliance on individual subjective judgments.
The annotations primarily include both \emph{structure level} components and \emph{thematic function level} components, as shown in Fig.~\ref{fig:SonataStructure}, with auxiliary labels such as onset noise and instrumentation.

\section{The \emph{Sonalyzer-Moz} framework\label{sec:method}}

The framework of the proposed \emph{Sonalyzer-Moz} is illustrated in Fig.~\ref{fig:Sonalyze}. It primarily consists of three core modules: Feature Embedding, Feature Aggregation, and Boundary Classification. 
This section describes each module in detail.

The input to the \emph{Sonalyzer-Moz} framework is the mel-spectrogram. 
To compute the mel-spectrogram \( M \), we set the sampling rate \( sr \) to 22,050~Hz and the hop length \( hop \) to 512 frames, and use 128 Mel filters. 
The mel-spectrogram value at frame \( x_i \) for the \( m \)-th Mel filter is computed as follows:
\[
M( x_i, m) = \sum_{freq=1}^{F} H_m(freq) \cdot S( x_i, freq),
\]
where \( S( x_i, freq) \) denotes the power spectrogram at frame \(  x_i \) and frequency bin \( freq \), and \( H_m(freq) \) represents the response of the \( m \)-th Mel filter at frequency bin \(freq \).
Here, \(F\) is the number of frequency bins resulting from the Short Term Fourier Transform.
In addition, the frame index $x_i \in \mathbb{Z}^{+}$ ranges from $1$ to $l^f = \left\lfloor Dur \cdot sr/hop \right\rfloor$, where $Dur$ is the duration of the input music, in seconds.

Next, the mel-spectrogram \(M\) is passed through the Feature Embedding Module, which consists of three convolutional (CNN) layers. 
The output shapes of these layers are shown in the corresponding part of Fig.~\ref{fig:Sonalyze}.
Each CNN layer is followed by a pooling layer: the first two use max pooling, while the final layer uses adaptive pooling to normalize the feature map size.
After the embedding module, the input mel-spectrogram \(M\), initially with one channel, is transformed into  \(C\) channels representation, denoted as $M'$.

Subsequently, \(M'\) will be aggregated. 
Specifically, the \(l^f\) frames' features will be sliced using a window of size \( \gamma \). Hence, the \( l^f \) frames will be segmented into \( \lfloor l^f/\gamma \rfloor \) slices. 
Subsequently, we perform feature reshaping, so that the output of the Feature Aggregation module is \( \sigma \in \mathbb{R}^{\lfloor l^f/\gamma \rfloor \times (\gamma \times C)} \). This step is crucial for enabling the \emph{Sonalyzer-Moz} framework to focus on upper-level features for structural analysis.

After feature aggregation, we compute a single-cycle position value for each aggregated frame index, denoted as \( pos_i \). The position is calculated as:
\[
pos_i = \frac{2\pi i}{\lfloor l^f / \gamma \rfloor - 1}, \quad i = 0, 1, \cdots, \lfloor l^f / \gamma \rfloor - 1.
\]
Each position is then encoded using sine and cosine functions as \( \sin(pos_i) \) and \( \cos(pos_i) \), respectively.
Additionally, we compute the dynamic self-similarity matrix (DSSM) based on $\sigma$ using cosine similarity, denoted as $DSSM \in [0,1]$. 
In a self-similarity matrix, larger values indicate higher similarity, and vice versa. 
In contrast, boundaries usually correspond to transitions between different components and therefore exhibit lower similarity. To emphasize these boundary regions, we consider $1 - DSSM$. Based on this dynamic matrix $1 - DSSM$, we further compute statistical values for each frame, including the sum and mean, followed by min-max normalization. Finally, the positional information and the statistical values are then concatenated with the aggregated feature \( \sigma \), resulting in \( \sigma' \in \mathbb{R}^{\lfloor l^f/\gamma \rfloor \times (\gamma \times C + 4)} \).

Before performing the classification on $\sigma'$, we apply $L_{lstm}$ layers to learn the local contextual features, so that the \emph{Sonalyzer-Moz} framework can better capture the transition features at each boundary. 
Finally, a multilayer perceptron (MLP) is used to perform binary classification, determining whether each frame corresponds to a structural boundary or not. 
After prediction, the frame indices are converted to specific timestamps for calculating the hit rate metric. 
This post-processing procedure is commonly used in music structure analysis and involves three steps: applying a sigmoid activation, detecting local maxima, and performing window-based filtering.



\section{Experiment Settings\label{sec:exp}}
For our experiments, we split the \emph{SoSA-Moz} dataset into training, validation, and test sets with a ratio of 8:1:1.
In addition, since \emph{SoSA-Moz} is a multi-version dataset, we carefully ensured that all versions of the same piece were assigned to the same subset to prevent data leakage.
To reduce randomness and ensure reproducibility, we fixed the random seed to 42 and set $\gamma$ to correspond to the number of frames in one second (i.e., $sr/hop\_length$). 
Consequently, every one-second segment of music features is aggregated within the \emph{Sonalyzer-Moz} framework. 
All experiments were conducted on an HPC platform equipped with an NVIDIA A100 GPU. 
Additionally, our experiments are conducted using CUDA 11.7 with PyTorch 2.0.0. 
The released \emph{SoSA-Moz} materials and the \emph{Sonalyzer-Moz} source code are publicly available on Figshare\protect\footnotemark[1].

To validate the proposed \emph{Sonalyzer-Moz} framework, we perform 
hyperparameter tuning, ablation studies, and comparative experiments.
Although this is the first work on \emph{sonata} form analysis based on audio files and combining deep learning techniques, we acknowledge that many structural analysis models have been developed for pop music. 
Therefore, we selected several recent works for pop music structure analysis as baselines for comparisons. 
The evaluation primarily focuses on HR3R, HR3P, and HR3F, which correspond to the recall, precision, and F1 score of the hit rate within a tolerance of $\pm$3 seconds~\cite{kim2023all, marmoret2023barwise}.
\begin{table}[!t]
\caption{Results for different hyperparameter values. Here, only the top 5 results are recorded.\label{tab:abl}}
\vspace{-3mm}

\centering
\begin{tabular}{c|ccc}

\toprule
\midrule
 ($C$, $h_{lstm}$, $L_{lstm}$)   & HR3P ($\%$)   & HR3R ($\%$) & HR3F ($\%$) \\\midrule
 (10, 1024, 5)   & \textbf{76.47}  &\textbf{77.17}&\textbf{76.24}  \\ 
(15, 256, 5)     & 68.46     & 78.44  &   72.38      \\
 (10, 512, 5)   & 67.04   &79.22& 71.97              \\
 (15, 512, 3)   & 62.92   &76.50& 68.47              \\
 (10, 1024, 3)   & 59.99   &77.22& 66.33              \\
\midrule \bottomrule

\end{tabular}
\vspace{-5mm}
\end{table}
\section{Experiment Results}

\subsection{Hyperparameter Tuning}
For hyperparameter tuning, we selected several key parameters and specified candidate values for each: the feature aggregation window, the number of feature channels in the embedding $C \in \{5, 10, 15\}$, the hidden size of the LSTM layers $h_{lstm} \in \{256, 512, 1024, 2048\}$, and the number of LSTM layers $L_{lstm} \in \{1, 3, 5\}$.
After performing a grid search, we record the results in Table~\ref{tab:abl}. 
The best-performing configuration (i.e., the \emph{champion}) has $C = 10$, $h^{\mathrm{lstm}} = 1024$, and $L_{lstm} = 5$, achieving a \textbf{76.24\%} HR3F. This indicates that after abstracting and compressing the aggregated features to 10 dimensions, a 5-layer LSTM with hidden size 1024 can effectively learn the boundary features.

\subsection{Ablation Experiment}

Under the \emph{champion} configuration, we conducted an ablation study, as shown in the row ``Feature Agg.'' of Table~\ref{table:abl}. 
First, when feature aggregation was omitted, we adjusted the post-processing thresholds accordingly to prevent excessive false positives caused by the small local maxima window. Nevertheless, the resulting HR3F of 30.38\% clearly confirms the importance of feature aggregation in the \emph{Sonalyzer-Moz} framework. This indicates that, in structure analysis, particularly for upper-level structures (i.e., EDR) in the \SON~form, the model needs to avoid focusing excessively on fine-grained details. In other words, local music features alone cannot provide reliable cues for boundary detection. Notably, this finding is consistent with musicological principles, where structural analysis typically relies on the overall content of each component rather than isolated local features.

Furthermore, our ablation experiments reveal that position encoding also plays a crucial role, as reported in the ``Single-cycle PE'' row of Table~\ref{table:abl}. Without positional information, model performance dropped by approximately 20\% HR3F, achieving only 56.72\% HR3F. In fact, for upper-level structures (i.e., EDR), although there is no explicit rule or prescribed proportion for the distribution of the three components within a piece, the significance of position encoding may suggest that, in actual musical compositions, there exists a certain regularity in the relative proportions of EDR components.
In addition, the results of the ablation study indicate that the advantage of LSTM in learning temporal sequences is indispensable for structure analysis. We also conducted a detailed evaluation of the dynamic self-similarity matrix (DSSM) component in the \emph{Sonalyzer-Moz} framework. As shown in Table~\ref{table:abl}, computing statistical features (i.e., sum and mean) from the dynamic self-similarity matrix provides effective guidance for the model to learn boundary characteristics.
\begin{table}[!t]
\caption{Ablation study of the \emph{Sonalyzer-Moz} framework.\label{table:abl}}
\vspace{-3mm}

\centering
\begin{tabular}{l|ccc}

\toprule
\midrule
  -w/o (exclusion)    & HR3P (\%)  & HR3R (\%) &  HR3F (\%)  \\ \midrule
  Feature Agg.   & 45.81      & 23.22& 30.38 \\ 
  Single-cycle PE    &    54.67  &  64.06 &  56.72    \\
DSSM &63.21   & 59.78&  58.58\\
LSTM  &   53.20  & 71.00 & 60.03        \\
SSM$^{sum}$ &73.02   & 55.56&  62.05\\
SSM$^{mean}$ &54.31     & 78.44&  63.27\\

  \midrule 
\textbf{\emph{Sonalyzer-Moz}}& \textbf{76.47}  &\textbf{77.17}&\textbf{76.24} \\  \midrule \bottomrule
\end{tabular}
\end{table}

\subsection{Comparison against conventional methods}
In this section, we compare our proposed Sonalyzer-Moz  against recent and representative methods for popular music, including the unsupervised method CBM~\cite{marmoret2023barwise} and supervised neural networks such as AllInOne~\cite{kim2023all} and SongFormer~\cite{hao2025songformer}. 
Although it is expected that methods designed for popular music are unlikely to be directly applicable to \emph{sonata} form analysis due to the significant differences in structural characteristics, this section not only presents the comparison results but also justifies the proposed architecture.

Specifically, CBM~\cite{marmoret2023barwise} is an unsupervised method that operates on mel-spectrogram features. Its block-enhanced self-similarity computation at the bar level serves a function similar to the feature aggregation proposed in \emph{Sonalyzer-Moz}. However, as an unsupervised method, CBM cannot benefit from fine-tuning. The CBM algorithm based on cosine similarity and covariance auto-similarity (i.e., row ``$\text{CBM}^{cos}$'' and row ``$\text{CBM}^{con}$'') achieves only 13.03\% HR3F and 8.93\% HR3F respectively, which indicates that this approach is largely unsuitable for \emph{sonata} form analysis.
In fact, CBM~\cite{marmoret2023barwise} is essentially a bar-level dynamic programming method. In the upper-level structure of the \emph{sonata} form (i.e., EDR), each component contains many musical phrases, which often have clear boundaries that may not align with the EDR-level boundaries. Consequently, bar-level retrieval in CBM is easily disrupted by these phrase-level boundaries. Therefore, even with feature aggregation, current unsupervised methods remain inadequate for \emph{sonata} form analysis. This limitation is one of the main motivations for our investigation into deep learning methods and the development of the proposed dataset.

\begin{table}[!t]

\caption{Comparison of \emph{Sonalyzer-Moz} with existing representative methods.\label{table:cp_sosa}}
\vspace{-3mm}

\centering
\begin{tabular}{l|ccc}

\toprule
\midrule
  -w/o (exclusion)    & HR3P (\%) & HR3R (\%)  & HR3F (\%)    \\ \midrule
  CBM$^{con}$~\cite{marmoret2023barwise}         & 5.57  & 23.84 & 8.93    \\
CBM$^{cos}$~\cite{marmoret2023barwise}         & 9.21  & 23.84 & 13.03 \\
SongFormer~\cite{hao2025songformer} &  11.30   &46.56& 18.12\\ 
AllInOne~\cite{kim2023all}   &  34.99   &75.11& 46.49\\

 \midrule 
\textbf{\emph{Sonalyzer-Moz}}& \textbf{76.47}  &\textbf{77.17}&\textbf{76.24} \\  \midrule \bottomrule

\end{tabular}
\vspace{-3mm}
\end{table}

On the other hand, AllInOne~\cite{kim2023all} is a supervised neural network model based on dilated attention, capable of learning interactions between different instruments in music. 
In addition, the model is trained jointly on multiple tasks, including structure segmentation and structure labeling, to enhance its overall effectiveness. 
However, at the EDR level in \emph{sonata} form, structures typically appear only in the form of EDR or EDRDR. Such structures do not provide sufficient training potential for the structure labeling task, and can often be accomplished using simple algorithms. Therefore, we only utilized the structure segmentation branch of AllInOne~\cite{kim2023all}.
Furthermore, another key design of AllInOne~\cite{kim2023all} is its instrument segmentation. In general, popular music involves multiple instruments, whereas the \emph{sonata} form we currently focus on primarily features solo piano or string instruments. While we still perform instrument differentiation, this strategy offers little benefit for solo piano music. After retraining the model on our dataset, the performance reached only 46.49\% HR3F (as reported in Table~\ref{table:cp_sosa}). 
Given that we have adapted AllInOne for our setting, this result cannot provide a comprehensive reflection of its performance in the context of \emph{sonata} form. However, the outcome can indicate that the architectures based on dilated attention and transformers do not exhibit significant effectiveness for \emph{sonata} form analysis.

Similar to AllInOne~\cite{kim2023all}, SongFormer~\cite{hao2025songformer} is also a transformer-based method. However, unlike AllInOne, SongFormer does not perform instrument separation. 
Instead, it segments the music representation into chunks of varying lengths, including 30-second and 420-second segments, to facilitate the learning of upper-level features. 
Nonetheless, the results reported in Table~\ref{table:cp_sosa} indicate that this approach performs worse than AllInOne.
Therefore, we can infer that although the 30-second and 420-second chunks may seemingly provide features at different levels, the use of non-overlapping consecutive chunks may disrupt the musical continuity. Furthermore, considering that both AllInOne and SongFormer incorporate transformer layers, albeit with different attention mechanisms, their suboptimal performance suggests that transformer architectures are currently unable to capture the most informative trainable features in the \emph{sonata} form structure.

Another point to note is that both AllInOne and SongFormer are trained jointly on structure segmentation and structure labeling tasks. In our evaluation, we only used the structure segmentation branch, which implies that structure labeling may play a non-negligible role in structural analysis for non-classical music. However, in the upper-level structure of the \emph{sonata} form (i.e., EDR level), structure labeling does not provide meaningful trainable information. Consequently, effective \emph{sonata} form analysis requires more targeted model design.

Overall, our \emph{Sonalyzer-Moz} model, although essentially relying only on the CNN layer and LSTM layer with a specific design, remains the most effective method for analyzing the EDR-level structures of \emph{sonata} form. We can further infer that future improvements in \emph{sonata} form analysis should focus on the model design and framework rather than simply increasing the model size (e.g., using larger transformers). Overall, our experiments demonstrate that \emph{Sonalyzer-Moz} can serve as an effective baseline method.

\section{Conclusion and Future Work\label{sec:conclusion}}

The \emph{SoSA-Moz} dataset proposed in this paper is the first large-scale auido-based \emph{sonata} form dataset with multi-level annotations, including both structure level and thematic function level labels. It enables and supports structure analysis via data driven approaches. 
Built on the \emph{SoSA-Moz} dataset, the proposed \emph{Sonalyzer-Moz} framework is the baseline model for \emph{sonata} form \emph{structure level} analysis. It not only demonstrates the effectiveness of neural network–based approaches for \emph{sonata} form analysis but also validates the capability and value of the \emph{SoSA-Moz} dataset in supporting such methods. As a baseline method for \emph{sonata} form analysis, although it currently achieves only 76.24\% F1 score at the structure level within a 
$\pm3$s error tolerance, comparisons with state-of-the-art methods developed for popular music show that the \emph{Sonalyzer-Moz} framework remains the most effective approach.

Based on our analysis, we recommend that future research focus on designing frameworks tailored to the musical features of \emph{sonata} form, rather than simply increasing model complexity or parameter count. 
We envision that this work will draw more attention to the field of classical music form structure analysis. Furthermore, we aim to inspire future studies to advance the analysis of \emph{sonata} comprehensively, addressing both the \emph{structure level} and the \emph{thematic function level} of this form, which is characterized by a typical hierarchical organization.

\section*{Acknowledgment}

This work was supported in part by the Advanced Computing Platform at Monash University Malaysia.

\printbibliography

\end{document}